\documentclass{LMCS}

\usepackage{color}
\usepackage{amssymb}
\usepackage{enumerate}

\usepackage{amsmath}
\usepackage{amsfonts}
\usepackage{611}
\usepackage{url}
\usepackage{hyperref}

\newtheorem{definition}[thm]{Definition}

\newtheorem{corollary}[thm]{Corollary}

\newtheorem{lemma}[thm]{Lemma}

\newcommand\iffl{\ensuremath{\leftrightarrow}}

\renewcommand\int{\mathbf{Z}}
\newcommand\rimpl{\ensuremath{\to}}

\newcommand{\ignore}[1]{}
\newcommand{\ov}[1]{\ensuremath{\overline{#1}}}
\newcommand{\nat}{\ensuremath{\mathbb{N}}}

\newcommand\reals{\ensuremath{\Vdash}}
\newcommand{\p}{\proves}

\newcommand{\g}{\Gamma}
\newcommand{\gp}{\Gamma \proves}

\newcommand{\sr}[1]{\SB{#1}_\rho}

\newcommand{\izfax}[1]{{\bf #1}}

\newcommand{\bool}{{bool}}
\renewcommand{\prop}{{prop}}
\newcommand{\ST}{*}
\newcommand{\st}{*}
\newcommand{\pr}[1]{\mathcal{#1}}
\newcommand{\Prop}{\ensuremath{P(1)}}

\def\doi{4 (3:5) 2008}
\lmcsheading%
{\doi}
{1--17}
{}
{}
{Jan.~31, 2007}
{Sep.~\phantom{0}9, 2008}
{}   

\begin{document}

\title{Extracting Programs from Constructive HOL Proofs via IZF
Set-Theoretic Semantics}

\author[R.L.~Constable]{Robert L. Constable\rsuper a}
\address{{\lsuper{a,b}}Department of Computer Science, Cornell University, Ithaca, NY,14853, USA}
\email{\{rc,wojtek\}@cs.cornell.edu}

\author[W.~Moczyd\l owski]{Wojciech Moczyd\l owski\rsuper b}
%\address{Department of Computer Science, Cornell University, Ithaca, NY,14853, USA}
%\email{wojtek@cs.cornell.edu}
\thanks{{\lsuper{a,b}}The authors have been partly supported by NSF grants DUE-0333526 and 0430161.}

\titlecomment{}

\begin{abstract} 
\noindent Church's Higher Order Logic is a basis for influential proof
assistants --- HOL and PVS.  Church's logic has a simple set-theoretic
semantics, making it trustworthy and extensible.  We factor HOL into a
constructive core plus axioms of excluded middle and choice. We
similarly factor standard set theory, ZFC, into a constructive core,
IZF, and axioms of excluded middle and choice. Then we provide the
standard set-theoretic semantics in such a way that the constructive
core of HOL is mapped into IZF.  We use the disjunction, numerical
existence and term existence properties of IZF to provide a program 
extraction capability from proofs in the constructive core.

We can implement the disjunction and numerical existence properties in two
different ways: one using Rathjen's realizability for IZF and the
other using a new direct weak normalization result for IZF by Moczyd\l owski. 
The latter can also be used for the term existence property. 
\end{abstract}

\maketitle

\section{Introduction} 

Church's Higher-Order logic \cite{Chu40,Lei94c} has been remarkably successful
at capturing the intuitive reasoning of mathematicians.  It was
distilled from {\em Principia Mathematica}, and is sometimes called the
Simple Theory of Types based on that legacy.  It incorporates the $\lambda$
calculus as its notation for functions, including propositional functions,
thus interfacing well with computer science, where the $\lambda$ calculus is
fundamental. 

One of the reasons Higher-Order logic is successful is that its axiomatic
basis is very small, and it has a clean set-theoretic semantics at a low
level of the cummulative hierarchy of sets (up to $\omega + \omega$) and can
thus be formalized in a small fragment of ZFC set theory.
This means it interfaces well with standard mathematics and
provides a strong basis for trust.  Moreover, the set theory semantics
is the basis for many extensions of the core logic; for example, it is
straightforward to add arrays, recursive data types, and records to the logic. 

Church's theory is the logical basis of two of the most successful
interactive provers used in hardware and software verification, HOL
\cite{GM93} and PVS \cite{ORS92}.
This is due in part to the two characteristics mentioned above in addition
to its elegant automation based on Milner's tactic mechanism and its elegant
formulation in the ML metalanguage.

Until recently, one of the few drawbacks of HOL was that its logical base did
not allow a way to express a constructive subset of the logic.  This issue
was considered by Harrison for HOL-light \cite{Har96}, and recently
Berghofer implemented a constructive version of HOL in the Isabelle
implementation \cite{Ber04,BN02} in large part to enable the extraction of
programs from constructive proofs.  This raises the question of finding a
semantics for HOL that justifies this intuitively sound extraction.
 
The standard justification for program extraction is based on logics that
embedded extraction deeply into their semantics; this is the case for the
Calculus of Inductive Constructions (CIC) \cite{CP-M90,BC04}, Minlog
\cite{BBS98}, Computational Type Theory (CTT) \cite{ABCEKLM05,book} or the closely related
Intuitionistic Type Theory (ITT) \cite{ML82,NPS90}. The mechanism of
extraction is built deeply into logic and the provers based on it, e.g. Agda
\cite{ACN90} on ITT, Coq \cite{CoqManV8} on CIC, MetaPRL \cite{HNC+03} and
Nuprl \cite{ACEKL00} on CTT.

In this paper we show that there is a way to provide a clean set-theoretic
semantics for HOL and at the same time use it to semantically justify program extraction.
The idea is to first factor HOL into its constructive core, say Constructive
HOL, plus the axioms of excluded middle and choice.  The semantics for this
language can be given in ZFC set theory, and if that logic is factored into
its constructive core, called IZF, plus excluded middle and choice (choice
is sufficient to give excluded middle), then in the standard semantics, IZF
provides the semantics for Constructive HOL.  Moreover, we can base program
extraction on the IZF semantics.

The constructive content of IZF is not as transparent as in the constructive
set theory CZF of Aczel \cite{Acz78}, as he is able to
interpret CZF in Type Theory, while no such interpretation is known for
IZF. However, it is not possible to express the impredicative nature of
Higher-Order Logic in CZF.  Also, IZF is not as expressive as Howe's ZFC
\cite{Howe96,Howe98a} with inaccessible cardinals and computational
primitives, but this makes IZF a more standard theory.

Our semantics is appealing not only because it factors so elegantly, but
also because the computational issues and program extraction can be reduced
to the standard constructive properties of IZF --- the disjunction,
numerical existence and term existence properties. 

We can implement the disjunction and numerical existence properties in two
different ways: one using Rathjen's realizability for CZF \cite{Rat05},
recently extended to IZF \cite{rathjen2006}, and the other using a new direct weak normalization result for IZF
by Moczyd\l owski \cite{jacsl2006,jatrinac2006}. The latter can also be used for the term existence
property.

In this paper, we provide a set-theoretic semantics for HOL which has the 
following properties: 
 
\begin{enumerate}[$\bullet$] 
\item It is as simple as the standard semantics, presented in Gordon and
Melham's \cite{GM93}.
\item It works in constructive set-theory. 
\item It provides a semantical basis for program extraction.
\item It can be applied to the constructive version of HOL recently
implemented in Isabelle-HOL as a means of using constructive HOL proofs
as programs.
\end{enumerate} 
 
This paper is organized as follows. In section \ref{hol} we present a
version of HOL. In section \ref{semantics} we define set-theoretic
semantics.  Section \ref{izf} defines constructive set theory IZF and states
its main properties. We show how these properties can be used for program
extraction in section \ref{extraction}. 
 
\section{Higher-order logic}\label{hol}

In this section, we present in detail higher-order logic. 
There are two syntactic categories: \emph{terms} and \emph{types}. The types are 
generated by the following abstract grammar:
\[
\tau ::= nat\ |\ bool\ |\ \prop\ |\ \tau \to \tau\ |\ \tau \times \tau
\]
The distinction between $bool$ and $prop$ corresponds to the distinction
between the two-element type and the type of propositions in type theory, or
between the two-element object and the subobject classifier in category theory or,
as we shall see, between $2$ and the set of all subsets of $1$ in constructive set theory.

The terms of HOL are generated by the following abstract grammar:
\[
t ::= x_\tau\ |\ c_\tau \ |\ (t_{\tau \to \sigma}\ u_{\tau})_\sigma\ |\
(\lambda x_\tau.\ t_\sigma)_{\tau \to \sigma}\ |\ (t_\tau, s_\sigma)_{\tau
\times \sigma}
\]

Thus each term $t_\alpha$ in HOL is annotated with a type $\alpha$, which we
call \emph{the type of $t$}. We will often skip annotating of terms with
types, this practice should not lead to confusion, as the implicit type
system is very simple. Terms of type $\prop$ are called \emph{formulas}. 

The free variables of a term $t$ are denoted by $FV(t)$ and defined as
usual. We consider $\alpha$-equivalent terms equal. The notation $t[x:=u]$
stands for a capture-avoiding substitution and denotes the result of
substituting $u$ for $x$ in the term $t$. 

Our version of HOL has a set of built-in constants. To increase readability,
we write $c : \tau$ instead of $c_\tau$ to provide the information about the
type of $c$. If the type of a constant involves
$\alpha$, it is a constant \emph{schema}, there is one constant for each
type $\tau$ substituted for $\alpha$. There are thus constants $=_{bool}$,
$=_{nat}$ and so on. 
\[
\bot : prop \qquad  \top : prop \qquad =_{\alpha} : \alpha \times \alpha \to prop 
\]
\[
\rimpl : prop \times prop \to prop \qquad \land : prop \times prop \to prop \qquad \lor : prop
\times prop \to prop
\]
\[
\forall_\alpha : (\alpha \to prop) \to prop \qquad \exists_\alpha : (\alpha \to prop) \to prop \qquad \varepsilon_\alpha : (\alpha \to \prop) \to \alpha
\]
\[
0 : nat \qquad S : nat \to nat \qquad false : bool \qquad true : bool
\]

We present the proof rules for HOL in a sequent-based natural deduction
style. A \emph{sequent} is a pair $(\Gamma, t)$, where $\Gamma$ is a list of formulas
and $t$ is a formula. The free variables of a context are the free variables of
all its formulas. A sequent $(\g, t)$ is written as $\gp t$.
We write binary constants (equality, implication, etc.) using infix notation. We use standard abbreviations for
quantifiers: $\forall a : \tau.\ \phi$ abbreviates $\forall_\tau (\lambda a_\tau.\ \phi)$, 
similarly with $\exists a : \tau.\ \phi$. The proof rules for HOL are as
follows:
\[
\infer[t \in \g]{\gp t}{} \qquad \infer{\gp t = t}{} 
\qquad \infer[x_\tau \notin FV(\g)]{\gp \lambda x_\tau.\ t = \lambda x_\tau.\ s}{\gp t = s}
\]
\[
\infer{\gp t \land s}{\gp t & \gp s} \qquad \infer{\gp t}{\gp t \land s}
\qquad \infer{\gp s}{\gp t \land s} \qquad \infer{\gp \top}{}
\]
\[
\infer{\gp t \lor s}{\gp t} \qquad \infer{\gp t \lor s}{\gp s} \qquad
\infer{\gp u}{\gp t \lor s & \g, t \p u & \g, s \p u}
\]
\[
\infer{\g \p t \to s}{\g, t \p s} \qquad \infer{\gp t}{\gp s
\to t & \gp s}
\qquad \infer{\gp t[x:=s]}{\gp s = u & \gp t[x:=u]}
\]
\[
\infer{\gp \exists_\alpha(f_{\alpha \to prop})}{\gp f_{\alpha \to prop}\ t_\alpha}
\qquad
\infer[x_\alpha\ \mbox{new}]{\gp u}{\gp \exists_\alpha (f_{\alpha \to prop})
& \g, f_{\alpha \to prop}\ x_\alpha \p u}
\]
\newpage
Finally, we list HOL axioms.

\begin{enumerate}[(1)]
\item (FALSE) $\bot = \forall b : prop.\ b$.
\item (FALSENOTTRUE) $false = true \to \bot$. 
\item (BETA) $(\lambda x_\tau.\ t_\sigma) s_\tau = t_\sigma[x_{\tau}:=s_\tau]$.
\item (ETA) $(\lambda x_\tau.\ f_{\tau \to \sigma}\ x_\tau) = f_{\tau \to
\sigma}$, where $x \notin FV(f)$. 
\item (FORALL) $\forall_{\alpha} = \lambda P_{\alpha \to prop}.\ (P = \lambda x_\alpha. \ \top)$.
\item (P3) $\forall n : nat.\ (0 = S(n)) \to \bot$. 
\item (P4) $\forall n, m : nat.\ S(n) = S(m) \to n = m$.
\item (P5) $\forall P : nat \to prop.\ P(0) \land (\forall n : nat.\ P(n) \to P(S(n))) \to \forall n : nat.\ P(n)$.
\item (BOOL) $\forall x : bool.\ (x = false) \lor (x = true)$.
\item (EM) $\forall x : prop.\ (x = \bot) \lor (x = \top)$.
\item (CHOICE) $\forall P : \alpha \to \prop.\ \forall x : \alpha.\ P\ x \to P (\varepsilon_{(\alpha \to
prop) \to \alpha}(P))$. 
\end{enumerate}	

Our choice of rules and axioms is redundant. Propositional connectives,
for example, could be defined in terms of quantifiers and $bool$. 
However, we believe that this makes the account of the semantics clearer and
shows how easy it is to define a sound semantics for such system. Our
presentation is based on the core part of the theory of \cite{GM93}. It does
not include type definitions and parametric polymorphism. We believe extending it to
incorporate these features should not be very difficult. 

The theory CHOL (Constructive HOL) arises by taking away from HOL the axioms
(CHOICE) and (EM).

We write $\p_H \phi$ and $\p_C \phi$ to denote that HOL and CHOL,
respectively, proves $\phi$. We will generally use letters $\pr{P}, \pr{Q}$ to denote
proof trees. A notation $\pr{P} \p_C \phi$ means that $\pr{P}$ is a proof tree in CHOL
of $\phi$. 

\section{Semantics}\label{semantics}

\subsection{Set theory}

The set-theoretic semantics needs a small part of the cumulative hierarchy --- $R_{\omega + \omega}$ is
sufficient to carry out all the constructions. The Axiom of Choice is necessary
in order to define the meaning of the $\varepsilon$ constant. For this
purpose, $C$ will denote a\footnote{Note that if we want to pinpoint $C$, we need to assume more than AC, as the
existence of a definable choice function for $R_{\omega + \omega}$ is not
provable in ZFC.} necessarily non-constructive function such that for any $X, Y \in R_{\omega +
\omega}$:
\begin{enumerate}[$\bullet$]
\item If $X$ is non-empty, then $C(X, Y) \in X$.
\item If $X$ is empty and $Y$ is non-empty, then $C(X, Y) \in Y$. 
\item Otherwise, $C(X, Y)$ is $\emptyset$. 
\end{enumerate}

Recall that in the world of set theory, $0 = \emptyset$, $1 = \{ 0 \}$ and $2 = \{ 0, 1 \}$.
Classically $\Prop$, the set of all subsets of $1$, is equal to $2$. This is not the case constructively; there is no
uniform way of transforming an arbitrary subset of $1$ into an element
of $2$. In fact, it is easy to see that $P(1) = 2$ entails the law of excluded
middle:
\begin{lemma}
If $\Prop = 2$, then for any $\phi$, $\phi$ or $\lnot \phi$.
\end{lemma}
\proof
Suppose $\Prop = 2$ and take a formula $\phi$. Consider $A = \{ x \in 1\ |\ \phi \}$ and $B
= \{ x \in 1\ |\ \lnot \phi \}$. Since $A \cup B \in \Prop$, $A \cup B \in 2$,
so either $A \cup B = 0$ or $A \cup B = 1$. In the former case, $0 \notin A$
and $0 \notin B$. Then we have $\lnot \phi$ because from $\phi$ we obtain $0 \in
A$, which is a contradiction. But we also have $\lnot \lnot \phi$ because from $\lnot
\phi$ we obtain $0 \in B$, which is also a contradiction. Thus we have refuted the assumption $
A \cup B=0$, so $A \cup B = 1$. Therefore $0 \in A \cup B$, so either $0 \in A$ in which case $\phi$, or $0 \in B$ in which case $\lnot \phi$. So either $\phi$ or
$\lnot \phi$.\qed

The following helpful lemma, however, does hold in a constructive world:
\begin{lemma}\label{lp01}
If $A \in P(1)$, then $A = 1$ iff $0 \in A$. 
\end{lemma}
\ignore
{
\proof
Left-to-right direction is immediate, for right-to-left we have $A \subseteq
1$ and need to show that $1 \subseteq A$. Suppose $B \in 1$, then $B = 0$,
but $0 \in A$, so $B \in A$.\qed
}
Let us also define precisely the function application operation in set theory. We borrow the definition from 
\cite{Acz99a}. 
\[
App(f, x) = \{ z\ |\ \exists y.\ z \in y \land (x, y) \in f \}
\]
The advantage of using this definition over an intuitive one (``the unique $y$ such that $(x, y) \in f$'') is 
that it is defined for all sets $f$ and $x$. Partiality of $App$ would
entail serious problems in the constructive setting.
This definition is equivalent to the standard one when $f$ is a function:
\begin{lemma}
If $f$ is a function from $A$ to $B$ and $x \in A$, then $App(f, x)$ is the
unique $y$ such that $(x, y) \in f$. 
\end{lemma}
\proof
Let $y$ be the unique element of $B$ such that $(x, y) \in f$.
If $z \in App(f, x)$ then there is $y'$ such that $z \in y'$ and $(x, y')
\in f$. Since $y' = y$, $z \in y$. For the other direction, if $z \in y$,
then obviously $z \in App(f, x)$.\qed

From now on, the notation $f(x)$ means $App(f, x)$. We will also use a lambda
notation in set theory to define functions: $\lambda x \in A.\ B(x)$ means $\{ (x, B(x)) \ | \ x \in A \}$.

\subsection{The definition of the semantics}

We first define a meaning $\SB{\tau}$ of a type $\tau$ by structural induction on $\tau$.
\begin{enumerate}[$\bullet$]
\item $\SB{nat} = \nat$. 
\item $\SB{bool} = 2$. 
\item $\SB{prop} = \Prop$. 
\item $\SB{\tau \times \sigma} = \SB{\tau} \times \SB{\sigma}$, where $A \times
B$ denotes the cartesian product of sets $A$ and $B$. 
\item $\SB{\tau_1 \to \tau_2} = \SB{\tau_1} \to \SB{\tau_2}$, where $A \to
B$ denotes the set of all functions from $A$ to $B$. 
\end{enumerate}

The meaning of a constant $c_\alpha$ is denoted by $\SB{c_\alpha}$ and is 
defined as follows.

\begin{enumerate}[$\bullet$]
\item $\SB{=_\alpha} = \lambda (x_1, x_2) \in \SB{\alpha}\times \SB{\alpha}.\ \{ x \in 1\ | \ x_1 = x_2 \}$.
\item $\SB{\to} = \lambda (b_1, b_2) \in \SB{prop} \times \SB{prop}.\ \{ x \in 1\ | \ x \in b_1 \to x \in b_2 \}$.
\item $\SB{\lor} = \lambda (b_1, b_2) \in \SB{prop} \times \SB{prop}.\ b_1 \cup b_2$. 
\item $\SB{\land} = \lambda (b_1, b_2) \in \SB{prop} \times \SB{prop}. \ b_1 \cap b_2$. 
\item $\SB{false} = \SB{\bot} = 0$. 
\item $\SB{true} = \SB{\top} = 1$. 
\item $\SB{\forall_\alpha} = \lambda f \in \SB{\alpha} \to \SB{prop}.\ \bigcap_{a \in \SB{\alpha}} f(a)$.
\item $\SB{\exists_\alpha} = \lambda f \in \SB{\alpha} \to \SB{prop}.\ \bigcup_{a \in \SB{\alpha}} f(a)$.
\item $\SB{\varepsilon_\alpha} = \lambda P \in \SB{\alpha} \to \SB{\prop}.\
C(P^{-1}(\{1\}), \SB{\alpha})$.
\item $\SB{0} = 0$.
\item $\SB{S} = \lambda n \in \nat.\ n + 1$
\end{enumerate}

Standard semantics, presented for example by Gordon and Melham in \cite{GM93}, uses a
truth table approach --- implication $\phi \to \psi$ is false iff
$\phi$ is true and $\psi$ is false etc. It is easy to see that with excluded
middle, our semantics is equivalent to the standard one.

\begin{lemma}[ZF]
For any $A, B \in P(1)$, $\SB{\to}(A, B)= 0$ iff $A = 1$ and $B = 0$. 
\end{lemma}
\proof
Suppose $\SB{\to}(A, B) = 0$. Then $\{ x \in 1\ | \ x \in A \to x \in B \} =
0$, so $0 \notin \{ x \in 1\ | \ x \in A \to x \in B \}$, so it is not the
case that $0 \in A \to 0 \in B$, so $0 \in A$ and $0 \notin B$. Thus, $A =
1$ and $B = 0$. The other direction is easy.\qed

The definition of our semantics is not original. The meaning of logical constants is essentially a
combination of the fact that any complete lattice with pseudo-complements is a model for higher-order
logic and that $P(1)$ is a complete lattice with pseudo-complement defined
in the clause for $\rimpl$ \cite{Rasiowa}. Similar semantics for HOL have also been provided in 
category-theoretical setting \cite{LS86}. The novelty of our approach lies
in \emph{utilizing} this kind of semantics for the purpose of program extraction in
Section \ref{extraction}. 

To present the rest of the semantics, we need to introduce environments. An
\emph{environment} is a function from HOL variables to sets such that
$\rho(x_\tau) \in \SB{\tau}$. We will use the symbol $\rho$ exclusively for environments.
The meaning $\SB{t}_\rho$ of a term $t$ is parameterized by an environment
$\rho$ and defined by structural induction on $t$:

\begin{enumerate}[$\bullet$]
\item $\SB{c_\tau}_\rho = \SB{c_\tau}$. 
\item $\SB{x_\tau}_\rho = \rho(x_\tau)$.
\item $\SB{s\ u}_\rho = App(\SB{s}_\rho, \SB{u}_\rho)$.
\item $\SB{\lambda x_\tau.\ u}_\rho = \{ (a, \SB{u}_{\rho[x_\tau:=a]})\ |\ a \in
\SB{\tau} \}$.
\item $\SB{(s, u)}_\rho = (\SB{s}_\rho, \SB{u}_\rho)$. 
\end{enumerate}

\subsection{The properties of the semantics}

There are several standard properties of the semantics we have defined. 

\begin{lemma}[Substitution Lemma]
For any terms $t, s$ and environments $\rho$,  $\SB{t}_{\rho[x:=\SB{s}_\rho]} = \SB{t[x:=s]}_\rho$. 
\end{lemma}
\proof
By structural induction on $t$. Case $t$ of:
\begin{enumerate}[$\bullet$]
\item $c$ --- the claim is obvious. 
\item $x$. Then $\SB{x}_{\rho[x:=\SB{s}_\rho]} = \SB{s}_\rho =
\SB{x[x:=s]}_\rho$. 
\item $u\ v$. Then $\SB{u\ v}_{\rho[x:=\SB{s}]} =
App(\SB{u}_{\rho[x:=\SB{s}_\rho]}, \SB{v}_{\rho[x:=\SB{s}_\rho]})$. By the inductive
hypothesis, this is equal to $App(\SB{u[x:=s]}_\rho, \SB{v[x:=s]}_\rho) = \SB{u[x:=s]\
v[x:=s]}_\rho = \SB{t[x:=s]}_\rho$. 
\item $(u, v)$. Similar to the previous case. 
\item $\lambda y_\tau.\ u$. Without loss of generality we may assume that $y
\notin \{ x \} \cup FV(s)$. Then $\SB{t}_{\rho[x:=s]} = \{ (a,
\SB{u}_{\rho[x:=\SB{s}_\rho][y:=a]})\ |\ a \in \SB{\tau} \}$. By the inductive
hypothesis, this is equal to $\{ (a, \SB{u[x:=s]}_{\rho[y:=a]})\ |\ a \in
\SB{\tau} \} = \SB{(\lambda y_\tau.\ u[x:=s])}_\rho = \SB{t[x:=s]}_\rho$.\qed
\end{enumerate}

\begin{lemma}\label{typesnotempty}
For any type $\alpha$, $\exists x.\ x \in \SB{\alpha}$.
\end{lemma}
\proof
Easy.\qed

\begin{lemma}\label{fvfv}
If $x_\sigma \notin FV(t)$, then for any $b \in \SB{\sigma}$, $\sr{t} = \SB{t}_{\rho[x_\sigma:=b]}$.
\end{lemma}
\proof
Straightforward induction on $t$. We only show the case when $t = \lambda
y_\tau.\ u$. Without loss of generality we can assume that $y \neq x$. 
We have $\sr{t} = \{ (a, \SB{u}_{\rho[y:=a]})\ |\ a \in \SB{\tau} \}$. Since
$x \notin FV(u)$, by the inductive hypothesis this is equal to
$\{ (a, \SB{u}_{\rho[y:=a][x:=b]})\ |\ a \in \SB{\tau} \}$. Since $x \neq
y$, this is also equal to $\{ (a, \SB{u}_{\rho[x:=b][y:=a]})\ |\ a \in
\SB{\tau} \} = \SB{\lambda y_\tau.\ u}_{\rho[x:=b]}$.\qed

\begin{lemma}\label{lin}
For any $\rho$, $\SB{t_\alpha}_\rho \in \SB{\alpha}$. 
\end{lemma}
By induction on $t$. Case $t$ of:
\begin{enumerate}[$\bullet$]
\item $x_\tau$. The claim follows by the definition of environments. 
\item $c_\tau$. We proceed by case analysis of $c$. We show the interesting
cases. 
\begin{enumerate}[$-$]
\item $\forall_\alpha$. The type of $c$ is $(\alpha \to prop) \to prop$.
We need to show that if $f$ is a function from $\SB{\alpha}$ to $\Prop$,
then $\bigcap_{a \in \SB{\alpha}} f(a)$ is in $\Prop$. Since for any $a \in
\SB{\alpha}$, $f(a) \in \Prop$ and $\Prop$ is closed under intersections,
the claim follows.
\item $\exists_\alpha$. The proof is similar and follows by the fact that $\Prop$ is closed
under unions.
\item $\varepsilon_\alpha$. The type of $\varepsilon_\alpha$ is $(\alpha \to prop) \to
\alpha$. Take any function $F$ from $\SB{\alpha}$ to $\Prop$. Then $F^{-1}(\{1\})
\subseteq \SB{\alpha}$. By the definition of $C$, if $F^{-1}(\{1\}) \neq
\emptyset$, then $\SB{\varepsilon_\alpha}(F) \in \SB{\alpha}$. So suppose 
$F^{-1}(\{1\}) = \emptyset$. By Lemma \ref{typesnotempty}, $\SB{\alpha}$ is
not empty, so by the definition of $C$, $\SB{\varepsilon_\alpha}(F) \in \SB{\alpha}$ as well. 
\end{enumerate}
\end{enumerate}

In particular, this implies that for any formula $t$, $\sr{t} \subseteq 1$. 
So if we want to prove that $\sr{t} = 1$, then by Lemma \ref{lp01} it
suffices to show that $0 \in \sr{t}$.

\subsection{Soundness}

The soundness theorem establishes validity of the proof rules and axioms with respect to the semantics.

\begin{definition}
We write $\SB{\g}_\rho = 1$ if $\SB{t_1}_\rho = 1, {\ldots} , 
\SB{t_n}_\rho = 1$,  where $\g = t_1, t_2, {\ldots} , t_n$.
\end{definition}

\begin{thm}[Soundness]
If $\g \p t$ then for any $\rho$, if $\SB{\g}_\rho = 1$, then $\sr{t} = 1$.
\end{thm}
\proof
Straightforward induction on $\gp t$. We show several interesting cases. 
\begin{enumerate}[$\bullet$]
\item 
\[
\infer[t \in \g]{\gp t}{}
\]
The claim is trivial.
\item 
\[
\infer{\gp \lambda x_\tau.\ t = \lambda x_\tau.\ s}{\gp t = s}
\]
We need to show that $\{ (a, \SB{t}_{\rho[x_\tau:=a]})\ |\ a \in \SB{\tau} \} =
\{ (a, \SB{s}_{\rho[x_\tau:=a]})\ |\ a \in \SB{\tau} \}$. That is, that for any $a
\in \SB{\tau}$, $\SB{t}_{\rho[x_\tau:=a]} = \SB{s}_{\rho[x_\tau:=a]}$. Let $\rho' =
\rho[x_\tau:=a]$. We get the claim by the inductive hypothesis. 
\ignore{
\item  
\[
\infer{\gp t \land s}{\gp t & \gp s}
\]
Suppose $\sr{\g} = 1$. By IH, $0 \in \sr{t}$ and $0 \in \sr{s}$, so $0 \in
\sr{t} \cap \sr{s}$. 
\item
\[
\infer{\gp t}{\gp t \land s} \qquad \infer{\gp s}{\gp t \land s}
\]
Reverse the previous case to get the claims.
}
\ignore
{
\item 
\[
\infer{\gp t \lor s}{\gp t} \qquad \infer{\gp t \lor s}{\gp s} \qquad \infer{\gp u}{\gp t \lor s & \g, t \p u & \g, s \p u}
\]
The first two cases are easy. For the last one, suppose $\sr{\g} = 1$. By
IH, we know that $0 \in \sr{t} \cup \sr{s}$, so either $0 \in \sr{t}$ or $0
\in \sr{s}$. In both cases, by the rest of IH, $0 \in \sr{u}$, so we get the
claim.
}	
\item 
\[
\infer{\g \p t \rimpl s}{\g, t \p s}
\]
Suppose $\SB{\g}_\rho = 1$. We need to show that $0 \in \{ x \in 1\ |\
x \in \SB{t}_\rho \to x \in \sr{s} \}$. Since $0 \in 1$, assume $0 \in
\sr{t}$. Then $\SB{\g, t}_\rho = 1$. By the inductive hypothesis
$\sr{s} = 1$ thus also $0 \in \sr{s}$. 
\item 
\[
\infer{\gp s}{\gp t \to s & \gp t}
\]
Suppose $\sr{\g} = 1$. By the inductive hypothesis, $0 \in \{ x \in 1\ |\ x \in
\sr{t} \to x \in \sr{s} \}$ and $0 \in \sr{t}$, so easily $0 \in \sr{s}$.
\item 
\[
\infer{\gp t[x:=s]}{\gp s = u & \gp t[x:=u]}
\]
Assume $\sr{\g} = 1$. By the inductive hypothesis, $\sr{s} = \sr{u}$ and $\sr{t[x:=u]} = 1$. 
Using the Substitution Lemma we get $\sr{t[x:=u]} = \SB{t}_{\rho[x:=\sr{u}]} =
\SB{t}_{\rho[x:=\sr{s}]} = \sr{t[x:=s]}$.
\item 
\[
\infer{\gp \exists_\alpha(f_{\alpha \to prop})}{\gp f\ t_\alpha}
\]
Assume $\sr{\g} = 1$. We have to show that $0 \in \bigcup_{a \in
\SB{\alpha}} (\sr{f}(a))$, so that there is $a \in \SB{\alpha}$ such that $0
\in \sr{f}(a)$. By Lemma \ref{lin}, $\sr{t_\alpha} \in \SB{\alpha}$, so taking $a
= \sr{t_\alpha}$ we get the claim by the inductive hypothesis. 
\item 
\[
\infer[x_\alpha\ \mbox{new}]{\gp u}{\gp \exists_\alpha (f_{\alpha \to prop}) & \g, f\ x_\alpha \p u}
\]
Suppose $\sr{\g} = 1$. By the inductive hypothesis, there is $a \in \SB{\alpha}$ such that $0 \in
\sr{f}(a)$. Let $\rho' = \rho[x_\alpha := a]$. By the inductive hypothesis we get $0 \in
\SB{u}_{\rho'}$. As $x_\alpha\ \notin FV(u)$, by Lemma \ref{fvfv} $\sr{u} = 1$.\qed
\end{enumerate}

Having verified the soundness of the HOL proof rules, we proceed to verify
the soundness of the axioms.

\begin{thm}\label{soundness}
For any axiom $t$ of HOL and any $\rho$ defined on $FV(t)$,  $0 \in \sr{t}$.
\end{thm}
\proof
We proceed axiom by axiom and sketch the respective proofs.
\begin{enumerate}[$\bullet$]
\item (FALSE) $\sr{\bot} = \emptyset = \bigcap_{a \in \Prop} a = \sr{\forall b :
prop.\ b}$. The second equality follows by $0 \in \Prop$. 
\item (BETA) 
We have $\sr{(\lambda x_\tau.\ t_\sigma)\ s_\tau} =
App(\sr{\lambda x_\tau.\ t_\sigma}, \sr{s_\tau}) = 
App(\{ (a, \SB{t}_{\rho[x:=a]})\ |\ a \in \SB{\tau} \}, \sr{s_\tau}) =
\SB{t}_{\rho[x_\tau:=\sr{s_\tau}]} = $ (by the Substitution Lemma) $ =
\sr{t_\sigma[x_\tau:=s_\tau]}$.
\item (ETA) 
$\sr{\lambda x_\tau.\ f_{\tau \to \sigma} x_\tau} =
\{ (a, \SB{f\ x_\tau}_{\rho[x_\tau:=a]})\ |\ a \in \SB{\tau} \} = 
\{ (a, App(\SB{f}_{\rho[x_\tau:=a]}, a))\ |\ a \in \SB{\tau} \} = $ (since
$x_\tau \notin FV(f)$) $ = \{ (a, \sr{f}(a))\ |\ a \in \SB{\tau} \} =
\sr{f}$, as by Lemma \ref{lin}, $\sr{f} \in \SB{\tau} \to \SB{\sigma}$ and functions in
set theory are represented by their graphs.
\item (FORALL) We have:
\[
\sr{\forall_\alpha} = \{ (F, \bigcap_{a \in \SB{\alpha}}
F(a))\ |\ F \in \SB{\alpha} \to \Prop \}
\]
Furthermore:
\[
\quad \qquad \sr{\lambda F_{\alpha \to
prop}.\ F = \lambda x_\alpha. \ \top} = \{ (F, \{ z \in 1\ |\ F = \lambda x \in
\SB{\alpha}.\ 1 \})\ |\ F \in \SB{\alpha} \to \Prop \}
\]
So take any $F \in \SB{\alpha} \to \Prop$. It suffices to show that $\bigcap_{a
\in \SB{\alpha}} F(a) = \{ z \in 1\ |\ F = \lambda x \in \SB{\alpha}.\ 1 \}$.
We have $x \in \bigcap_{a \in \SB{\alpha}} F(a)$ iff for all $a \in
\SB{\alpha}$, $x \in F(a)$ and $x = 0$. This happens if and only if $x = 0$ and for all $a \in
\SB{\alpha}$, $F(a) = 1$ which is equivalent to $x \in \{ z \in 1\ |\ P =
\lambda x \in \SB{\alpha}.\ 1 \}$. The claim follows. 
\item The axioms $P3, P4, P5$ follow by the fact that natural numbers
satisfy the respective Peano axioms.
\item (BOOL) We need to show that $\sr{\forall_\bool.\ (\lambda x_\bool.\ x = false \lor
x = true)} = 1$. Unwinding the definition, this is equivalent to
$\bigcap_{x \in 2} (\{ z \in 1\ |\ x = 0 \} \cup \{ z \in 1\ |\ x = 1 \})
= 1$. and furthermore to: for all $x \in 2$ and $y$, $y \in \{ z \in 1\ |\ x = 0 \} \cup
\{ z \in 1\ |\ x = 1 \}$ iff $y = 0$. Take any $x \in 2$ and $y$. The
left-to-right direction is obvious, for the right-to-left direction, either $x = 0$ or
$x = 1$. In the former case, $0 \in \{ z \in 1\ |\ x = 0 \}$, in the latter $0 \in \{ z \in 1\ |\ x = 1 \}$.
\item (EM) We need to show that $\SB{\forall_\prop.\ (\lambda x_\prop.\ x = \bot \lor
x = \top)}_\rho = 1$. Reasoning as in the case of (BOOL), we find that this is
equivalent to: for all $x \in P(1)$ and $y$, $y \in \{ z \in 1\ |\ x = 0 \} \cup
\{ z \in 1\ |\ x = 1 \}$ iff $y = 0$. Suppose $x \in P(1)$. At this point, it is
impossible to proceed further constructively, all we know is that $x$ is a
subset of $1$, which does not provide enough information to decide whether $x =
0$ or $x = 1$. However, classically, using the rule of excluded middle, $P(1) = 2$ and we
proceed as in the previous case.
\item (CHOICE) 
We argue classically, so in particular $P(1) = 2$. We need to show that:
\[
\begin{array}{ll}
\SB{\forall_{\alpha \to \prop} (\lambda P_{\alpha \to \prop}.\
\forall_\alpha (\lambda x_\alpha.\ P x \rimpl P (\varepsilon_{(\alpha \to
\prop) \to \alpha}(P))} = 1, &\mbox{which is equivalent to}\\
\bigcap_{P \in \SB{\alpha} \to
2} \SB{\forall_\alpha (\lambda x_\alpha.\ P x \rimpl
P(\varepsilon_{(\alpha \to \prop) \to \alpha}(P))} = 1, & \mbox{which is
equivalent to}\\
\bigcap_{P \in \SB{\alpha} \to 2} \bigcap_{x \in \SB{\alpha}} \SB{
P x \rimpl P (\varepsilon_{(\alpha \to \prop) \to \alpha}(P))} =
1, &\mbox{which is equivalent to}\\
\end{array}
\]
\[
\bigcap_{P \in \SB{\alpha} \to 2} \bigcap_{x \in \SB{\alpha}} \{ a \in 1\ |\ a \in P(x) \to
a \in P(C(P^{-1}(\{1\}), \SB{\alpha})) \} = 1.
\]
To show this, it
suffices to show that for all $P \in \SB{\alpha} \to 2$, for all $x \in
\SB{\alpha}$, if $0 \in P(x)$ then $0 \in P(C(P^{-1}(\{1\}), \SB{\alpha}))$. Take any
$P$ and $x$. Suppose $0 \in P(x)$. Then $P(x) = 1$, so $x \in P^{-1}(\{1\})$.
Therefore $C(P^{-1}(\{1\}), \SB{\alpha})) \in P^{-1}(\{1\})$, so 
$P(C(P^{-1}(\{1\}), \SB{\alpha}) = 1$, which shows the claim.\qed
\end{enumerate}

\begin{corollary}
HOL is consistent: it is not the case that $\p_H \bot$.
\end{corollary}
\proof
Otherwise we would have $\SB{\bot} = \SB{\top}$, that is $0 = 1$.\qed

\section{IZF}\label{izf}

\begin{figure}[t]\label{f1}
\begin{enumerate}[$\bullet$]
\item \izfax{Extensionality} Two sets are equal if they have the same elements. 
\item \izfax{Empty Set} There is an empty set.
\item \izfax{Pairing} For any sets $a, b$, there is a set consisting of $a$ and $b$.
\item \izfax{Infinity} There is a set closed under the successor operation
and containing the empty set. 
\item \izfax{Union} For any set $a$, there is a set $\bigcup a$ which is a union of all elements of $a$.
\item \izfax{Power Set} For any set $a$, there is a set of all subsets of $a$.
\item \izfax{Separation} For any formula $\phi$, for any set $a$, there is a set of all elements of
$a$ satisfying $\phi$. 
\item \izfax{Replacement} For any formula $\phi(x, y, \ov{z})$, for any set $a$, if for all $x \in a$ there is
exactly one $y$ such that $\phi(x, y, \ov{z})$ holds, then there is a set $b$ such that
for all $x \in a$ there is $y \in b$ such that $\phi(x, y, \ov{z})$ holds.
\item \izfax{$\in$-Induction} For any formula $\phi(a, \ov{z})$, if for all sets $b$ 
$(\forall x \in b. \phi(x, \ov{z}))$ implies $\phi(b, \ov{z})$, then for all
$a$, $\phi(a, \ov{z})$ holds. 
\end{enumerate}
\caption{The axioms of IZF with Replacement}
\end{figure}

The essential advantage of the semantics in the previous section over
a standard one is that for the constructive part of HOL this semantics can be
defined in constructive set theory IZF.

An obvious approach to creating a constructive version of ZFC set theory is
to replace the underlying first-order logic with intuitionistic first-order
logic. As many authors have explained \cite{Myhill73,Bee85,McCarty86,Sce85}, the
ZF axioms need to be reformulated so that they do not imply the law of excluded
middle.  

In a nutshell, to get IZF from ZFC, the Axiom of Choice and Excluded Middle
are taken away and Foundation is reformulated as $\in$-induction. The axioms of IZF are thus Extensionality, Union, Infinity, 
Power Set, Separation, Replacement or Collection\footnote{There is a
difference, in particular the version with Collection does not satisfy Term
Existence Property (TEP), defined on the next page. A concerned reader can replace IZF with IZF$_R$ whenever TEP is used. }
and $\in$-Induction. The list of axioms for the version with Replacement can be found in
Figure~1. 
A detailed account of the theory can be found for
example in Friedman \cite{Fri73}. Besoon's book \cite{Bee85} and
\v{S}\v{c}edrov's paper \cite{Sce85} contain a lot of information on metamathematical properties of IZF and related set theories. 
For convenience, we assume that the first-order logic has built-in bounded quantifiers
($\forall x \in a.\ \phi$ and $\exists x \in a.\ \phi$), defined as
abbreviations in the standard way. We also include in the signature all the
set terms corresponding to the axioms of IZF --- $\nat, \bigcup t, P(a)$ etc.
For the full list, see \cite{jalmcs07}.

Myhill \cite{Myhill73} have proved several important properties of IZF:
\begin{enumerate}[$\bullet$]
\item Disjunction Property (DP) : If IZF $\p \phi \lor \psi$, then IZF $\p
\phi$ or IZF $\p \psi$.
\item Numerical Existence Property (NEP) : If IZF $\p \exists x \in \nat.\
\phi(x)$, then there is a natural number $n$ such that IZF $\p
\phi(\ov{n})$, where $\ov{n} = S(S({\ldots}(0)))$ and $S(x) = x \cup \{ x
\}$. 
\item Term Existence Property (TEP) : If IZF $\p \exists x.\ \phi(x)$, then
for some term $t$, IZF $\p \phi(t)$. 
\end{enumerate}

Moreover, the semantics and the soundness theorem for CHOL work in IZF, as
neither Choice nor Excluded Middle are necessary to carry out these
developments. Note that the existence of \Prop\ is crucial for the semantics. 

All the properties are constructive --- there is a
recursive procedure extracting a natural number, a disjunct or a term from a proof.
A trivial one is to look through all the proofs for the correct one. For
example, if IZF $\p \phi \lor \psi$, a procedure could enumerate all theorems
of IZF looking for either $\phi$ or $\psi$; its termination would be ensured by DP. 
We discuss more efficient alternatives in section \ref{dpnep}.

\section{Extraction}\label{extraction}

We will show that the semantics we have defined can serve as a basis for
program extraction from proofs. All that is necessary for program extraction
from constructive HOL proofs is provided by the semantics and the soundness
proof. Therefore, if one wants to provide an extraction mechanism for the
constructive part of the logic, it may be sufficient to carefully define set-theoretic semantics,
prove the soundness theorem and the extraction mechanism for IZF would take
care of the rest. We speculate on practical uses of this approach in section
\ref{conclusion}.

\subsection{IZF Extraction}

We first describe extraction from IZF proofs. To facilitate the description,
we will use a very simple fragment of type theory, which we call $TT^0$. 

The \emph{types} of $TT^0$ are generated by the following abstract grammar.
They should not be confused with HOL types; the context will make it clear
which types we refer to. 
\[
\tau ::= \ST\ |\ P_\phi\ |\ nat\ |\ bool\ |\ \tau \times \tau\ |\ \tau + \tau\ |\ \tau \to \tau
\]

We associate with each type $\tau$ of $TT^0$ a set of its elements, which are
finitistic objects. The set of elements of $\tau$ is denoted by $El(\tau)$
and defined by structural induction on $\tau$:

\begin{enumerate}[$\bullet$]
\item $El(\ST) = \{ \st \}$.
\item $El(P_\phi)$ is the set of all IZF proofs of the formula $\phi$.
\item $El(nat) = \nat$, the set of natural numbers.
\item $El(bool) = \{ true, false \}$.
\item $El(\tau_1 \times \tau_2) = El(\tau_1) \times El(\tau_2)$.
\item $M \in El(\tau_1 + \tau_2)$ iff either $M = inl(M_1)$ and $M_1 \in
El(\tau_1)$ or $M = inr(M_1)$ and $M_1 \in El(\tau_2)$.
\item $M \in El(\tau_1 \to \tau_2)$ iff $M$ is a method which given any
element of $El(\tau_1)$ returns an element of $El(\tau_2)$.
\end{enumerate}

In the last clause, we use an abstract notion of ``method''. It will not be
necessary to formalize this notion, but for the interested reader, 
all ``methods`` we use are functions provably recursive in $ZF + Con(ZF)$,
where $Con(ZF)$ denotes consistency of ZF. 

The notation $M : \tau$ stands for $M \in El(\tau)$. 

We call a $TT^0$ type \emph{pure} if it does not contain $\ST$ and $P_\phi$. There is a natural mapping of pure types $TT^0$ to sets. It is so similar to the
meaning of the HOL types that we will use the same notation.  
\begin{enumerate}[$\bullet$]
\item $\SB{nat} = \nat$.
\item $\SB{bool} = 2$.
\item $\SB{\tau \times \sigma} = \SB{\tau} \times \SB{\sigma}$.
\item $\SB{\tau + \sigma} = \SB{\tau} + \SB{\sigma}$, the disjoint union of
$\SB{\tau}$ and $\SB{\sigma}$. 
\item $\SB{\tau \to \sigma} = \SB{\tau} \to \SB{\sigma}$. 
\end{enumerate}
If a set (and a corresponding IZF term) is in a codomain of the map above, we
call it \emph{type-like}. If a set $A$ is type-like, then there is a unique
pure type $\tau$ such that $\SB{\tau} = A$. We denote this type $Type(A)$.
Thus, type-like sets are these ``generated'' by pure $TT^0$ types via
natural semantics. Formally, we define a recursive set $TL$ of IZF terms
such that for any $t \in TL$, $t$ is type-like and we can find effectively
$Type(A)$. The definition of $TL$ follows the definition above: $TL$ is the
smallest set such that $\nat, 2 \in TL$ and if $t, u \in TL$, then $t \times
u$, $t + u$ and $t \to u$ are also elements of $TL$. Thus, the sentence
``$A$ is type-like'' stands for ``$A \in TL$''. Note that for any term $t
\in TL$ we can find a term $t'$ such that IZF $\p t = t'$ and $t' \notin TL$
--- it suffices to take $t' \equiv t \cup \emptyset$. 

Before we proceed further, let us extend $TT^0$ with a new type $Q_\tau$,
where $\tau$ is any pure type of $TT^0$. Intuitively, $Q_\tau$ is the
provable counterpart of $\SB{\tau}$. Formally, the members of 
$El(Q_\tau)$ are pairs $(t, \pr{P})$ such that $\pr{P} \p_{IZF} t \in \SB{\tau}$ 
($\pr{P}$ is an IZF proof of $t \in \SB{\tau}$). 
Note that there is a natural mapping from closed HOL terms $M$ of type $\tau$ into
$Q_{\tau}$ --- it is easy to construct using Lemma \ref{lin} a proof $\pr{P}$ of
the fact that $\SB{M}_\rho \in \SB{\tau}$, so the pair $(\SB{M}_\rho, P) : Q_\tau$.
In particular, any natural number $n$ can be injected into $Q_{nat}$. The
set of pure types stays unchanged. 

We are going to tailor extraction from IZF proofs to the HOL logic. For this purpose, we
will specify which elements of IZF proofs/formulas carry interesting
computational content for us. We will use the type $\ST$ to mark the parts of
proofs we are not interested in.

We first define a helper function $T$, which takes a pure type $\tau$ and
returns another type. Intuitively, $T(\tau)$ is the type of the extract from
a statement $\exists x.\ x \in \SB{\tau}$. The function $T$ is defined by induction on $\tau$:
\begin{enumerate}[$\bullet$]
\item $T(bool) = bool$. 
\item $T(nat) = nat$.
\item $T(\tau \times \sigma) = T(\tau) \times T(\sigma)$. 
\item $T(\tau + \sigma) = T(\tau) + T(\sigma)$. 
\item $T(\tau \to \sigma) = Q_{\tau} \to T(\sigma)$. The rationale for this
definition is that in order to utilize an IZF function from $\SB{\tau}$ to $\SB{\sigma}$ we need to supply an element
of a set $\SB{\tau}$, which is an element of $Q_\tau$. 
\end{enumerate}

Furthermore, we assign to each formula $\phi$ of IZF a $TT^0$ type $\ov{\phi}$,
which intuitively describes the \emph{computational content} of an IZF proof of
$\phi$. We do it by induction on $\phi$:

\begin{enumerate}[$\bullet$]
\item $\ov{a \in b} = \ST$. 
\item $\ov{a = b} = \ST$ (atomic formulas carry no useful computational content). 
\item $\ov{\phi_1 \lor \phi_2} = \ov{\phi_1} + \ov{\phi_2}$.
\item $\ov{\phi_1 \land \phi_2} = \ov{\phi_1} \times \ov{\phi_2}$.
\item $\ov{\phi_1 \to \phi_2} = P_{\phi_1} \to \ov{\phi_2}$.
\item $\ov{\exists a \in A.\ \phi_1} = T(Type(A)) \times \ov{\phi_1}$, if $A$ is
type-like. 
\item $\ov{\exists a \in A.\ \phi_1} = \ST$, if $A$ is not type-like.
\item $\ov{\exists a.\ \phi_1} = \ST$.
\item $\ov{\forall a \in A.\ \phi_1} = Q_{Type(A)} \to \ov{\phi_1}$, if $A$ is type-like. 
\item $\ov{\forall a \in A.\ \phi_1} = \ST$, if $A$ is not type-like. 
\item $\ov{\forall a.\ \phi_1} = \ST$.
\end{enumerate}

The definition is tailored for HOL logic and could be extended to allow meaningful extraction from
a larger class of formulas. For example, we could extract a term from $\exists a.\
\phi_1$ using Term Existence Property. 

We present several natural examples of our translation in action:
\begin{enumerate}[(1)]
\item $\ov{\exists x \in \nat.\ x = x} = nat \times \st$.
\item $\ov{\forall x \in \nat.\  \exists y \in \nat.\ \phi} = Q_{nat} \to nat
\times \ov{\phi}$.
\item $\ov{\forall f \in \nat \to \nat.\  \exists x \in \nat.\ f(x) = 0} = 
Q_{nat \to nat} \to nat \times \st$. 
\end{enumerate}

These types are richer than what we intuitively would expect --- $nat$ in
the first case, $nat \to nat$ in the second and $(nat \to nat) \to nat$ in
the third, because any closed HOL term of type $nat$ or $nat \to nat$ can be
injected into $Q_{nat}$ or $Q_{nat \to nat}$ via the soundness theorem. The extra $\st$ can be easily
discarded from types (and extracts).

\begin{lemma}\label{ovsubst}
For any IZF term $t$, which is not type-like, $\ov{\phi[a:=\ov{t}]} = \ov{\phi}$. 
\end{lemma}
\proof
Straightforward induction on $\phi$.\qed

\begin{lemma}[IZF]\label{l2}
$(\exists a \in 2.\ \phi(a))$ iff $\phi(0) \lor \phi(1)$. 
\end{lemma}

We are now ready to describe the extraction function $E$, which takes an IZF
proof $\pr{P}$ of a
formula $\phi$ and returns an object of $TT^0$ type $\ov{\phi}$. We do it by
induction on $\phi$, checking on the way that the object returned is
of type $\ov{\phi}$. Recall that DP, TEP and NEP denote Disjunction, Term
and Numerical Existence Property, respectively. Case $\phi$ of:

\begin{enumerate}[$\bullet$]
\item $a \in b$ --- return $\st$. We have $\st : \ST$.
\item $a = b$ --- return $\st$. We have $\st : \ST$, too. 
\item $\phi_1 \lor \phi_2$. Apply DP to $\pr{P}$ to get a proof $\pr{P}_1$ of either
$\phi_1$ or $\phi_2$. In the former case return $inl(E(\pr{P}_1))$, in the latter
return $inr(E(\pr{P}_1))$. By the inductive hypothesis, $E(\pr{P}_1) : \ov{\phi_1}$ (or
$E(\pr{P}_1) : \ov{\phi_2}$), so $E(\pr{P}) : \ov{\phi}$ follows. 
\item $\phi_1 \land \phi_2$. Then there are proofs $\pr{P}_1$ and $\pr{P}_2$ such that
$\pr{P}_1 \p \phi_1$ and $\pr{P}_2 \p \phi_2$. Return a pair $(E(\pr{P}_1),
E(\pr{P}_2))$. By the inductive hypothesis, $E(\pr{P}_1) : \ov{\phi_1}$ and
$E(\pr{P}_2) : \ov{\phi_2}$, so
$(E(\pr{P}_1), E(\pr{P}_2)) : \ov{\phi_1 \land \phi_2}$. 
\item $\phi_1 \to \phi_2$. Return a function $G$ which takes an IZF proof
$\pr{Q}$ of $\phi_1$, applies $\pr{P}$ to $\pr{Q}$ (using the modus-ponens rule of the first-order logic) to get a proof
$\pr{R}$ of $\phi_2$ and returns $E(\pr{R})$. By the inductive hypothesis, any
such $E(\pr{R})$
is in $El(\ov{\phi_2})$, so $G : P_{\phi_1} \to \ov{\phi_2}$.
\item $\exists a \in A.\ \phi_1(a)$, where $A$ is type-like. Let $T = Type(A)$.
We proceed by induction on $T$. Case $T$ of:
\begin{enumerate}[$-$]
\item $bool$. By Lemma \ref{l2}, we have $\phi_1(0) \lor \phi_1(1)$. Apply DP to
get a proof $\pr{Q}$ of either $\phi_1(0)$ or $\phi_1(1)$. Let $b$ be $false$ or $true$,
respectively. Return a pair $(b, E(\pr{Q}))$. By the inductive hypothesis,
$E(\pr{Q}) :
\ov{\phi_1(\SB{b})}$. By Lemma \ref{ovsubst}, since $\sr{b}$ is not
type-like, $E(\pr{Q}) : \ov{\phi_1}$, so
$(b, E(Q)) : T(bool) \times \ov{\phi} = \ov{\exists a \in 2.\ \phi_1(a)}$.
\item $nat$. Apply NEP to $\pr{P}$ to get a natural number $n$ and a proof
$\pr{Q}$ of $\phi_1(\ov{n})$. Return a pair $(n, E(\pr{Q}))$. By the inductive hypothesis,
$E(\pr{Q}) : \ov{\phi_1(\ov{n})}$. By Lemma \ref{ovsubst}, since we can
assume without loss of generality that $\ov{n}$ is not type-like, $E(\pr{Q}) : \ov{\phi_1}$, so $(n, E(\pr{Q})) : T(nat)
\times \ov{\phi_1}$. 
\item $(\tau, \sigma)$. Construct a proof $\pr{Q}$ of $\exists a_1 \in \SB{\tau}
\exists a_2 \in \SB{\sigma}.\ a = <a_1, a_2> \land \phi_1$. Let $M = E(\pr{Q})$.
By the inductive hypothesis $M$ is a pair $<M_1, M_2>$ such that $M_1 : T(\tau)$
and $M_2 : \ov{\exists a_2 \in \SB{\sigma}.\ a = <a_1, a_2> \land \phi_1}$.
Therefore $M_2$ is a pair $<M_{21}, M_{22}>$, $M_{21} : T(\sigma)$ and
$M_{22} : \ov{a = <a_1, a_2> \land \phi_1}$. Therefore $M_{22}$ is a pair
$<N, O>$, where $O : \ov{\phi_1}$. Therefore $<M_1, M_{21}> : T(\tau \times
\sigma)$, so $<<M_1, M_{21}>, O> : T(\tau \times \sigma) \times \ov{\phi_1}$ and we
are justified to return $<<M_1, M_{21}>, O>$. 
\item $\tau + \sigma$. Construct a proof $\pr{Q}$ of $(\exists a \in
\SB{\tau}.\ \phi_1) \lor (\exists a \in \SB{\sigma}.\ \phi_1)$. Apply DP to get
the proof $\pr{Q}_1$ of (without loss of generality) $\exists a \in
\SB{\tau}.\ \phi_1$. Let $M = E(\pr{Q}_1)$. By the inductive hypothesis, $M = <M_1,
M_2>$, where $M_1 : T(\tau)$ and $M_2 : \ov{\phi_1}$. Return $<inl(M_1),
M_2>$, which is of type $(T(\tau + \sigma), \ov{\phi_1})$. 
\item $\tau \to \sigma$. Use TEP to get a term $f$ such that $(f \in
\SB{\tau} \to \SB{\sigma}) \land \phi_1(f)$. Construct proofs $\pr{Q}_1$ of $\forall
x \in \SB{\tau} \exists y \in \SB{\sigma}. f(x) = y$ and $\pr{Q}_2$ of
$\phi_1(f)$. Without loss of generality, we can assume that $f$ is not
type-like. By the inductive hypothesis and Lemma \ref{ovsubst}, $E(\pr{Q}_2) :
\ov{\phi}$. Let $G$ be a function which works as follows: $G$ takes a pair
$(t, \pr{R})$ such 
that $\pr{R} \p t \in \SB{\tau}$, applies $\pr{Q}_1$ to $t, \pr{R}$ to get a
proof $\pr{R}_1$ of $\exists y \in \SB{\sigma}.\ f(t) = y$ and calls
$E(\pr{R}_1)$ to get a term $M$. By the inductive
hypothesis, $M : \ov{\exists y \in \SB{\sigma}.\ f(t) = y}$, so $M = <M_1, M_2>$,
where $M_1 : T(\sigma)$. The function $G$ returns $M_1$. Our extraction procedure
$E(\pr{P})$ returns $<G, E(\pr{Q}_2)>$. The type of $<G, E(\pr{Q}_2>)$ is
$(\pr{Q}_\tau \to T(\sigma)) \times \ov{\phi_1}$ which is exactly $(T(\tau \to
\sigma)) \times \ov{\phi_1}$.
\end{enumerate}
\item $\exists a \in A.\ \phi_1(a)$, where $A$ is not type-like. Return $\st$. 
\item $\exists a.\ \phi_1(a)$. Return $\st$.
\item $\forall a \in A.\ \phi_1(a)$, where $A$ is type-like. Return a function $G$
which takes an element $(t, \pr{Q})$ of $Q_{Type(A)}$, applies $\pr{P}$ to
$t$ and $\pr{Q}$ to get a proof $R$ of $\phi_1(t)$, and returns $E(\pr{R})$. Without loss of
generality, we can assume that $t$ is not type-like. By the inductive hypothesis and
Lemma \ref{ovsubst}, $E(\pr{R}) : \ov{\phi_1}$, so $G : Q_{Type(A)} \to
\ov{\phi_1} = \ov{\forall a \in A.\ \phi_1(a)}$.
\item $\forall a \in A.\ \phi_1(a)$, where $A$ is not type-like. Return $\st$.
\item $\forall a.\ \phi_1(a)$. Return $\st$. 
\end{enumerate}

\subsection{HOL extraction}

As in case of IZF, we will show how to do extraction from a subclass of CHOL
proofs. The choice of the subclass is largely arbitrary, our choice
illustrates the method and can be easily extended.

We say that a CHOL formula is \emph{extractable} if it is generated by the
following abstract grammar, where $\tau$ varies over pure $TT^{0}$ types and $\oplus \in \{ \land, \lor,
\to \}$. 
\[
\phi ::= \forall x : \tau.\ \phi\ |\ \exists x : \tau.\ \phi\ |\ \phi \oplus \phi\
|\ \bot\ |\ t = t
\]
\newcommand{\ppp}{{\rho[a_1:=b_1,{\ldots} , a_n:=b_n]}}
\newcommand{\srr}[1]{\SB{#1}_{\rho'}}
We will define extraction for CHOL proofs of extractable formulas.
By~Theorem~\ref{soundness}, if CHOL $\p \phi$, then
IZF $\p 0 \in \SB{\phi}$. We need to slightly transform this IZF proof 
in order to come up with a valid input to $E$ from the previous section. To
this means, for any extractable $\phi(a_1, {\ldots} , a_n)$ we define an
IZF formula $\phi'(b_1, {\ldots} , b_n)$ such that IZF $\p 0 \in
\SB{\phi(a_1, {\ldots} , a_n)}_{\rho[a_1:=b_1, {\ldots} , a_n := b_n]} \iffl
\phi'$. The formula $\phi'$ is essentially $\phi$ with type membership information replaced by set membership information. We define $\phi'$ by induction on
$\phi$, checking the correctness on the way. We work in IZF. Let $\rho' =
\ppp$. Thus we want to show IZF $\p 0 \in \srr{\phi} \iffl \phi'$. Case $\phi$ of:
\begin{enumerate}[$\bullet$]
\item $\bot$. Take $\phi' \equiv 0 \in \srr{\bot}$. The correctness is trivial. 
\item $t = s$. Take $\phi' \equiv 0 \in \srr{t = s}$. The correctness is trivial. 
\item $\phi_1 \lor \phi_2$. By the inductive hypothesis we get $\phi_1'$ and
$\phi_2'$ such that $0 \in \srr{\phi_1} \iffl \phi_1'$ and $0 \in
\srr{\phi_2} \iffl \phi_2'$. Take $\phi' \equiv \phi_1' \lor \phi_2'$.
We have $0 \in \srr{\phi_1 \lor \phi_2}$ iff 
$0 \in \srr{\phi_1}$ or $0 \in \srr{\phi_2}$ iff $\phi_1' \lor \phi_2'$, which
shows the claim.
\item $\phi_1 \land \phi_2$. By the inductive hypothesis we get $\phi_1'$ and
$\phi_2'$ such that $0 \in \srr{\phi_1} \iffl \phi_1'$ and $0 \in \srr{\phi_2} \iffl \phi_2'$.
Set $\phi' \equiv  \phi_1' \land \phi_2'$. The correctness follows easily. 
\item $\phi_1 \to \phi_2$. By the inductive hypothesis we get $\phi_1'$
such that $0 \in \srr{\phi_1} \iffl \phi_1'$ and 
$\phi_2'$ such that $0 \in \srr{\phi_2} \iffl \phi_2'$. Set $\phi' = \phi_1'
\to \phi_2'$. The correctness follows easily. 
\item $\forall a : \tau.\ \phi_1(a, a_1, {\ldots} , a_n)$.
By the inductive hypothesis we get $\phi_1'(b, b_1, {\ldots} , b_n)$ such
that $\forall b, b_1, {\ldots} , b_n$, $0 \in \SB{\phi_1'}_{\rho'[a:=b]
} \iffl \phi_1'$. Set $\phi' \equiv \forall
a \in \SB{\tau}.\ \phi_1'(a, b_1, {\ldots} , b_n)$. For the correctness,
we have $0 \in \srr{\forall a : \tau.\
\phi_1(a, a_1, {\ldots} , a_n)}$ iff $\forall A \in \SB{\tau}$, $0 \in
\SB{\phi_1}_{\rho'[a:=A]}$. By the inductive
hypothesis, this is equivalent to $\forall A \in \SB{\tau}.\ \phi_1'(A,
b_1,{\ldots} , b_n)$ which is precisely $\phi_1'$. 
\item $\exists a : \tau.\ \phi_1$.
By the inductive hypothesis we get $\phi_1'(b, b_1, {\ldots} , b_n)$ such
that 
\[
\forall b, b_1, {\ldots} , b_n.\ 0 \in \SB{\phi_1'}_{\rho'[a:=b]} \iffl \phi_1.
\]
 Set $\phi' \equiv \exists a \in \SB{\tau}.\ \phi_1'(a, b_1, {\ldots} , b_n)$. The correctness follows
as in the previous case. 
\end{enumerate}

Now we can finally define the extraction process. Suppose CHOL $\p \phi$, where $\phi$
is closed and extractable. Let $\rho$ be the empty environment. Using the soundness theorem, construct an IZF proof $P$ that $0
\in \sr{\phi}$. Use the definition above to get $\phi'$ such that IZF $\p 0
\in \sr{\phi} \iffl \phi'$ and using $P$ obtain a proof $R$ of $\phi'$.
Finally, apply the extraction function $E$ to $R$ to get the computational extract.

\subsection{Implementation issues}\label{dpnep}

The extraction process is parameterized by the implementation of NEP, DP and
TEP for IZF. Obviously, searching through all IZF proofs to get a
witnessing natural number, term or a disjunct would not be a very effective method. 
We discuss two alternative approaches.

The first approach is based on realizability. Rathjen
defines a realizability relation in  \cite{Rat05} for weaker, predicative constructive set
theory CZF. For any CZF proof of a formula $\phi$, there is a realizer $e$ such
that the realizability relation $e \reals \phi$ holds, moreover, this
realizer can be found constructively from the proof. Realizers provide
the information for DP and NEP --- which of the disjuncts holds and the
witnessing number. They could be implemented using lambda terms. 
These results have been also recently extended to IZF \cite{rathjen2006}.
The approach has the drawback of not providing the proof of TEP, which
would restrict the extraction process from statements of the form $\exists x \in \SB{\tau}.\
\phi$ to atomic types $\tau$. Moreover, the gap between the existing theoretical
result and possible implementation is quite wide.

The second, more direct approach is based on Moczyd\l owski's proof in
\cite{jacsl2006} of weak
normalization of the lambda calculus $\lambda Z$ corresponding to proofs in IZF. The normalization is used to prove NEP, DP and TEP for the
theory and the necessary information is extracted from the normal form of the
lambda term corresponding to the IZF proof. Thus in order to provide the implementation of DP, NEP and 
TEP for IZF, it would suffice to implement $\lambda Z$, which is specified
completely in \cite{jacsl2006,jatrinac2006}.

An alternative approach has been presented by Berghofer \cite{Ber04}. He
defines extraction for a constructive variant of HOL logic directly in
the generic theorem prover Isabelle and uses realizability to justify its
correctness. His approach could likely be tailored to our CHOL, so that it
would yield extracts equivalent to ours. An exciting project would be to
\emph{formalize} IZF and both methods of extraction in Isabelle and show
their equivalence and correctness. 

\section{Conclusion}\label{conclusion}

We have presented a computational semantics for HOL via standard
interpretation in intuitionistic set theory. The semantics is clean, simple and 
agrees with the standard one. 

The advantage of this approach is that the extraction mechanism is
completely external to Constructive HOL. Using only the semantics, we can take 
any constructive HOL proof and extract from it computational information. 
No enrichment of the logic in normalizing proof terms is necessary. 

The separation of the extraction mechanism from the logic makes the logic
very easily extendable. For example, inductive datatypes and subtyping 
have clean set-theoretic semantics, so can easily be added to HOL preserving consistency, as witnessed in PVS. As the semantics
would work constructively, the extraction mechanisms from section \ref{extraction} could be easily adapted to incorporate
them. Similarly, one could define a set-theoretic semantics for the constructive version of 
HOL implemented in Isabelle (\cite{Ber04,BN02}) in the same spirit, with the
same advantages. 

The modularity of our approach and the fact that it is much easier to give
set-theoretic semantics for the logic than to prove normalization, could make 
the development of new trustworthy provers with extraction capabilities much
easier and faster.

We would like to thank anonymous reviewers for their helpful comments. 

\bibliography{rc}
\bibliographystyle{alpha}

\end{document}